\documentclass[prb,twocolumn,superscriptaddress,amsfonts,amssymb,showpacs]{revtex4}

\usepackage{graphicx}
\usepackage{dcolumn}
\usepackage{bm}
\usepackage{times}
\usepackage{amsmath}
\usepackage{color}

\newcolumntype{/}{D{/}{/}{2,2}}  
\newcolumntype{.}{D{.}{.}{0}}  

\begin{document}

\title{Insight into the temperature dependent properties of the ferromagnetic Kondo lattice YbNiSn}

\author{A. Generalov}
\affiliation{MAX-Laboratory, Lund University, Box 118, 22100 Lund, Sweden}

\author{D. A. Sokolov}
\affiliation{School of Physics and CSEC, Universuty of Edinburgh, Edinburgh EH9 3FD, UK}
\affiliation{Max-Planck-Institut f\"{u}r Chemische Physik fester Stoffe, D-01187 Dresden, Germany}

\author{A. Chikina}
\affiliation{Institut f\"{u}r Festk\"{o}rperphysik, Technische Universit\"{a}t Dresden, D-01062 Dresden, Germany}

\author{Yu. Kucherenko}
\affiliation{Institute for Metal Physics, National Academy of Science of Ukraine, UA-03142 Kiev, Ukraine}

\author{V. N. Antonov}
\affiliation{Institute for Metal Physics, National Academy of Science of Ukraine, UA-03142 Kiev, Ukraine}

\author{L. V. Bekenov}
\affiliation{Institute for Metal Physics, National Academy of Science of Ukraine, UA-03142 Kiev, Ukraine}

\author{S. Patil}
\affiliation{Department of Physics, Indian Institute of Technology, Banaras Hindu University, Varanasi-225001, India}

\author{A. D. Huxley}
\affiliation{School of Physics and CSEC, Universuty of Edinburgh, Edinburgh EH9 3FD, UK}

\author{J. W. Allen}
\affiliation{Randall Laboratory, University of Michigan, 450 Church St., Ann Arbor, MI 48109-1040, USA}

\author{K.~Matho}
\affiliation{Institut N\'{e}el, C.N.R.S. and Universit\'{e} Grenoble Alpes, BP 166, 38042 Grenoble cedex 9, France}

\author{K. Kummer}
\affiliation{European Synchrotron Radiation Facility, 71 Avenue des Martyrs, Grenoble, France}

\author{D. V. Vyalikh}
\affiliation{Saint Petersburg State University, Saint Petersburg 198504, Russia}
\affiliation{Donostia International Physics Center (DIPC), Departamento de Fisica de Materiales and CFM-MPC UPV/EHU, 20080 San Sebastian, Spain}
\affiliation{IKERBASQUE, Basque Foundation for Science, 48011 Bilbao, Spain}

\author{C. Laubschat}
\affiliation{Institut f\"{u}r Festk\"{o}rperphysik, Technische Universit\"{a}t Dresden, D-01062 Dresden, Germany}

\date{\today}

\begin{abstract}
Analyzing temperature dependent photoemission (PE) data of the ferromagnetic Kondo-lattice (KL) system YbNiSn in the light of the Periodic Anderson model (PAM) we show that the KL behavior is not limited to temperatures below a temperature $\overline{T}_K$, defined empirically from resistivity and specificic heat measurements. As characteristic for weakly hybridized Ce and Yb systems, the PE spectra reveal a 4$f$-derived Fermi level peak, which reflects contributions from the Kondo resonance and its crystal electric field (CEF) satellites. In YbNiSn this peak has an unusual temperature dependence: With decreasing temperature a steady linear increase of intensity is observed which extends over a large interval ranging from 100~K down to 1~K without showing any peculiarities in the region of $\overline{T}_K \sim T_C$= 5.6 K. In the light of the single-impurity Anderson model (SIAM) this intensity variation reflects a linear increase of 4$f$ occupancy with decreasing temperature, indicating an onset of Kondo screening at temperatures above 100~K. Within the PAM this phenomenon could be described by a non-Fermi liquid like $T$-\,linear damping of the self-energy which accounts phenomenologically for the feedback from the closely spaced CEF-states. 
  
\end{abstract}

\pacs{ }

\maketitle

\section{Introduction}

Intermetallic rare-earth (RE) systems based on Ce, Eu or Yb form a prototype of strongly correlated electron systems, where the interplay between almost localized $4f$ and itinerant valence states results in a wealth of extraordinary phenomena. Among them are ultra-heavy quasi-particle excitations (heavy fermions) out of Fermi-liquid (FL) or non-FL ground states that also compete with localized groundstates, showing magnetic order.~\cite{gegenwart2008,Coleman} At a quantum critical point (QCP), various ground states can become degenerate, which is typically achieved by applying pressure or changing the chemical composition.

In that regard, YbNiSn assumes a somewhat unique role. Already at ambient pressure and without chemical doping it shows a competition between the on-site exchange interaction that leads to formation of the Kondo lattice (KL) and the inter-site exchange interaction that leads to magnetic order. As a stoichiometric material with a low level of defects, YbNiSn broadens the range of experimental techniques available for studies of the QCP, such as quantum oscillations in the electrical resistivity and magnetization, but also photoemission (PE). Moreover, the KL competes in this compound with ferromagnetic (FM) order, not antiferromagnetic (AFM), as in most cases. 

In this work, we focus on the $T$-\,dependent properties in zero magnetic field, that were explored by means of electrical resistivity and specific heat measurements as well as PE spectroscopy. For simulations of the PE data, we have applied both the single-impurity Anderson model (SIAM) ~\cite{Anderson} and the periodic Anderson model (PAM). The modeling results are checked for their consistency. 

\begin{figure}[tbp!]
\begin{center}
  \includegraphics[width=0.80\columnwidth]{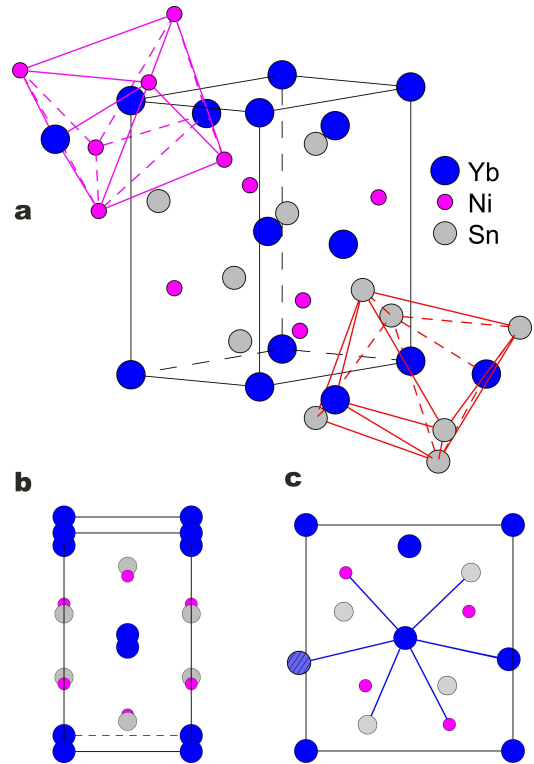}
\end{center}
\caption{(Color online) (a) Schematic representation of the YbNiSn crystal structure. Note, that the origin is shifted with respect to its symmetry group position to put the Yb atoms in the vertices of the unit cell for better visualization. (b) The unit cell projected on the $yz$ plane. The coordinate system is slightly rotated around $y$ axis in order to show all atoms in the layer. (c) The unit cell projected on the $xz$ plane. The bonds of the Yb atom within the $xz$ atomic layer are shown.}
\label{struc}
\end{figure}

The studied ternary compound YbNiSn crystallizes in the orthorhombic
$\varepsilon$-TiNiSi type structure with the space group symmetry $Pnma$
(number 62). The unit cell contains 4 formula units. All atoms are found in
layers parallel to the $xz$ plane and each layer contains all three
kinds of atoms. Crystallographic studies of the compound YbNiSn were
performed in Ref. \onlinecite{NNH+92} and yielded the unit cell parameters
$a=$6.960 \AA, $b=$4.410 \AA, $c=$7.607 \AA, at room temperature. Atoms are located at the
4$c$ Wyckoff positions: Yb at the (0.4897, 0.25, 0.2970), Ni at the
(0.2026, 0.25, 0.5857), and Sn at the (0.8074, 0.25, 0.5866) positions.
The nearest neighbor environment of the Yb atom consists of two distorted octahedra of Ni and Sn atoms, shown in Fig.~\ref{struc}. The interatomic distances for Yb-Ni are between 2.969 \AA\ ~and 3.266 \AA\ and for Yb-Sn between 3.068 \AA\ and 3.182 \AA. The next nearest neighbors of an Yb atom are four Yb atoms located at the vertices of a much distorted tetrahedron; two of them are at the distance of 3.552 \AA  ~(shown in the polyhedra in Fig.~\ref{struc}) and the other two at 3.797 \AA.

\section{Experiment}

Single crystals of YbNiSn were grown by melting stoichiometric amounts of the elements welded in Molybdenum crucibles and annealing at $800^\circ\text{C}$ for 120 hours. Powder X-ray diffraction confirmed the absence of impurity phases. Large single crystals produced by this method showed a small mosaic of $\sim$ 0.5 degree (FWHM) measured in the cold neutron beam and residual resistivity ratio (RRR) between 40 and 100 depending on direction of current with respect to the crystallographic orientation. Heat capacity was measured using Physical Property Measurements System by Quantum Design. Electrical resistivity was measured using a conventional four-probe technique with a lock-in amplifier with the current parallel to the $a$-axis.

PE experiments were performed at the $1^3$-ARPES instrument of beamline UE112~PGM-2 at BESSY-II synchrotron radiation facility and at the SIS instrument of the Swiss Light Source. Both instruments are equipped with a Scienta~R4000 photoelectron energy analyzer. The SIS instrument has a 6-axis CARVING manipulator with high angular precision which is ideally suited for ARPES experiments down to $10$~K. The $1^3$-ARPES instrument at BESSY-II allows ARPES measurements at temperatures close to $1$~K with ultra-high energy resolution. The samples were cleaved in ultra-high vacuum and explored by photoelectron spectroscopy. The temperature-dependent measurements were performed always going from high to low temperatures in order to avoid fast sample aging.

\section{Results of resistivity and specific heat measurements}

\begin{figure}[h]
\includegraphics[width=\linewidth]{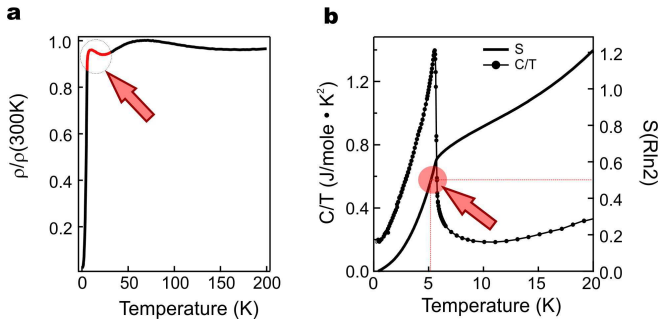}
\caption{\label{fig:epsart} Bulk properties of YbNiSn (magnetic field \textbf{H}\,=0): (a) $T$-\,dependence of the electrical resistivity along the \textit{a}-axis. The rise below 150~K signals the Kondo screening of a \textbf{\textit{J}}\,$=7/2$ multiplet. The broad feature at $\sim$70 K corresponds to the freezing out of the lowest CEF excitation. The KL formation sets in below the maximum at $\sim$10 K (arrow). The resistivity drop is accentuated by the suppression of spin disorder scattering in the FM phase. (b) $T$-\,dependence of the specific heat $C/T$ and the entropy $S(T)$. The large, lambda-like anomaly at $T_C$ = 5.6~K marks the FM transition. The arrow points to the fact that only half the entropy of a two level system, $S=\frac{1}{2}R$ln2, is recovered at $T_C$. The red dot indicates the zone of intense competition between FM order and KL formation.}
\label{Bulk Properties}
\end{figure}

The electrical resistivity $\rho(T)$ along the \textit{a}-axis in the orthorhombic crystal structure is shown in Fig.~\ref{Bulk Properties}a. A quantitative modeling, including strong anisotropy,~\cite{NNH+92} requires an orbitally degenerate KL with exchange parameters that are orbital- and momentum-dependent. The KL model has the same relation to the PAM as the single site Kondo model to the SIAM: They are linked via the  Schrieffer-Wolff transformation,~\cite{Cornut} valid when the state of Yb is almost trivalent.
 
The resistivity has two Kondo-like regimes, $d\rho/dT<0$. The higher one points to the presence of Kondo screening on a \textbf{\textit{J}}\,$=7/2$ multiplet of $f$-orbitals. Because of a large  spin-orbit (SO) splitting in Yb, the excited multiplet \textbf{\textit{J}}\,$=5/2$ is an example of degrees of freedom that are frozen out before any observable binding energy (BE) can be gained from the Kondo screening. In this case, perturbation theory is sufficient to define a Kondo temperature $T_K^{(8)}$ for an effective multiplicity N$_f$=2\,\textbf{\textit{J}}\,+1=8.~\cite{Cornut} Whether $T_K^{(8)}\,(k_B\equiv1)$ is the right energy scale for fluctuations within the  multiplet, depends further on the crystal-electric field (CEF) splittings. From inelastic neutron scattering,~\cite{adroja98-physb} the lowest excitation is placed  at $\sim$120~K and two additional doublets of the orthorhombic level scheme appear closely spaced. The plateau between 50 and 100~K, followed by the second regime with $d\rho/dT<0$, are non perturbative effects, attributed to the progressive freezing of excited CEF levels and the survival of an effective Kramers doublet. As the non crossing approximation SIAM\,+\,NCA reveals,~\cite{Kroha} the many body wave function of this ground doublet is dressed by the higher CEF orbitals, to the extent that they have contributed to the BE of the Kondo screening before being frozen out. The resistivity drop below $\sim$10~K is the coherent effect, captured only by a lattice model.
 
Specific heat and entropy measurements are shown in Fig.~\ref{Bulk Properties}b. The FM transition at  $T_C$ is marked by the lambda-like anomaly in $C/T$. Two features below $T_C$ provide evidence that Kondo screening continues in the FM state: (i) The low-$T$ limit of $C/T$, $\gamma\sim$200 mJ/mole$\cdot$K$^{2}$, is two orders of magnitude higher than that of an uncorrelated metal like Cu. An even higher value of $\sim$300~mJ/mole$\cdot$K$^{2}$ was reported previously for polycrystalline arc-molten samples.~\cite{kasaya92} This large $\gamma$ value indicates the formation of heavy quasi particles (QP). (ii) In a localized moment system, involving only N$_f$=2, the entropy recovered at the transition should approach $R$\,ln2, where $R$ is the gas constant. As working definition for a QP scale  $\overline{T}_K$ we take the point where $S(T)\sim\frac{1}{2}R$\,ln2.~\cite{klingner11-prb} The red dot indicates that $\overline{T}_K$  is not sharply defined. The range includes $T_C=$5.6~K but also a sharp negative peak at $\sim$ 7~K in the thermopower.~\cite{adroja98-physb} In the discussion of our PE simulations, the question arises whether this empirical scale can be related to the width of a QP band in the PAM, also denoted as $\overline{T}_K$.~\cite{Benlagra,notations} For the purpose of including orbital effects in the PAM, we define  $\overline{T}_K^{(N_f)}$, specifying the effective degeneracy. The respective scale in the SIAM is defined without the over-bar. The fact that $S(T)$ continues to rise smoothly beyond $R$\,ln2 signals again the necessity to take N$_f>2$ into account.

\section{Results from photoemission and discussion in light of Anderson models}

\subsection{SIAM versus PAM}

The multiple competitions, seen in the bulk properties, make YbNiSn very attractive for a PE study, in order to round off recent insights gained in YbRh$_2$Si$_2$ and YbIr$_2$Si$_2\,$.~\cite{vyal10,patil14,Yb-val,kummer2015}  Since PE covers low as well as high BE, the interpretations have to use either the SIAM or the PAM, without resorting to the Schrieffer-Wolff transformation. That one is dealing with KL physics has to be checked by the valence condition that the number of holes in the $4f$ shell is only slightly below $n_f=1$.

The SIAM interpretation of the "Kondo peak" in angle integrated PE, particularly its $T$-dependence, has been a lively topic ever since the 1980's,~\cite{Allen} in which SO and CEF splittings play a major role. It is instructive to briefly retrace some highlights, because they vindicate the opinion that PE is able to make contact with bulk properties. Patthey et al.~\cite{patthey} first applied SIAM+NCA to CeSi$_2$. Later, an extended and quasi-linear $T$-dependence was observed in this compound ~\cite{reinert,ehm} that turned out to be robust enough to survive an extrinsic convolution with a broad energy resolution function.~\cite{figure9} As discussed below, this may be relevant for our present observations on YbNiSn.
A resonance position slightly below $E_F$, as expected for Yb, was first checked on YbAl$_3$ by L. H. Tjeng et al,~\cite{tjeng93} where $T$-\,dependence in agreement with NCA was also confirmed. Applicability to both Ce and Yb contributed essentially to the general acceptance of the SIAM picture. The progress in energy resolution enabled Reinert et al. ~\cite{reinert,ehm} to resolve pairs of SO and CEF satellites on both sides of the Kondo resonance. The simulation of spectral weights and detailed $T$-\,dependence of satellite features with SIAM+NCA ~\cite{Kroha} confirms the ubiquitous presence of SO and CEF splittings.

The argument for a SIAM interpretation has always been that the angle integrated intensity is proportional to a density of states (DOS), i.e.: a trace over momentum resolved intensities.
The current challenge to PE is therefore to resolve features near $E_F$ that can be attributed to a QP-DOS. For example, the renormalized band theory for YbRh$_2$Si$_2$,~\cite{Zwicknagl} exhibits van Hove singularities (VHS) in the DOS that would be wiped out in the corresponding SIAM. Will it be possible for PE with extreme energy resolution to pinpoint evidence for strongly renormalized VHS?
Such new challenges require a better understanding of the subtle differences between SIAM and PAM, assuming identical parameters for the $f$-shell and the hybridization. Finally, for quantitative answers, the specific energy convolutions of PE as a one-step process must also be taken into account.~\cite{Krasovskii} 

The dynamical mean field theory (DMFT) offers a key to the general relationship and also the difference between SIAM and PAM. The SIAM provides a local "dynamical mean field" that solves a lattice self-energy (SE) for the PAM. Although this SE is "local", in the sense that it is independent of momentum, translational symmetry is recovered by using the proper Dyson equation for a lattice, which is block diagonal in the symmetry label $k$. The meaning of this label, which extends only over a single Brillouin zone (BZ), is  that of crystal momentum modulo umklapp. In order to model an angle integrated PE spectrum, the partial $i$-DOS for the set of orbital characters "$i$" are needed, together with matrix elements depending on the photon energy.  Intense theoretical studies exist only for the case of a Kramers doublet (N$_f$=2).~\cite{tahvildar,Pruschke,Meyer,Amaricci,Benlagra,Kainz}  In the paramagnetic state, the problem reduces to finding the Kramers degenerate $c$-DOS and $f$-DOS, without further orbital distinction. Van Hove singularities, broadened by FL damping, are noticeable in DMFT\,+\,NRG calculations on lattices in finite dimension and at low $T$.~\cite{Benlagra,Kainz} To reach the limit $T\rightarrow0$, the numerical renormalization group solution for the solver, SIAM+NRG,~\cite{Costi} bridges certain deficiencies of SIAM\,+\,NCA. The Anderson parameters ~\cite{Anderson} assumed in these studies are generally too small ($U$) or too large ($V$)  for direct application to Ce and Yb. 
In principle, the material specific LDA\,+\,DMFT modeling of compounds with strong correlations requires all orbitals of a partly filled $d$- or $f$- shell,~\cite{Georges} but the actual application to RE compounds poses an extreme numerical challenge. In the pioneering work on CeIrIn$_5$,~\cite{shim,choi} the CEF splittings are not yet included.~\cite{shim1}

The toy model, proposed below, is intended to provisorily fill this gap, until DMFT routines for the PAM with CEF splitting become available. It has the translationally invariant features of a PAM but introduces a simplified picture for the $f$-shell, with multiplicity N$_f$ treated as phenomenological parameter. The hierarchy of energy scales can be inferred, via the Schrieffer-Wolff transformation, from the SIAM.~\cite{Cornut} The toy model aims to explain at least some of the remaining discrepancies in earlier interpretations for Ce ~\cite{Joyce,Arko} or Yb ~\cite{Lawrence,Weibel,Blyth} compounds. 

\subsection{Photoemission spectra and the SIAM simulation}

\begin{figure}
\centering
\includegraphics[width=0.90\columnwidth]{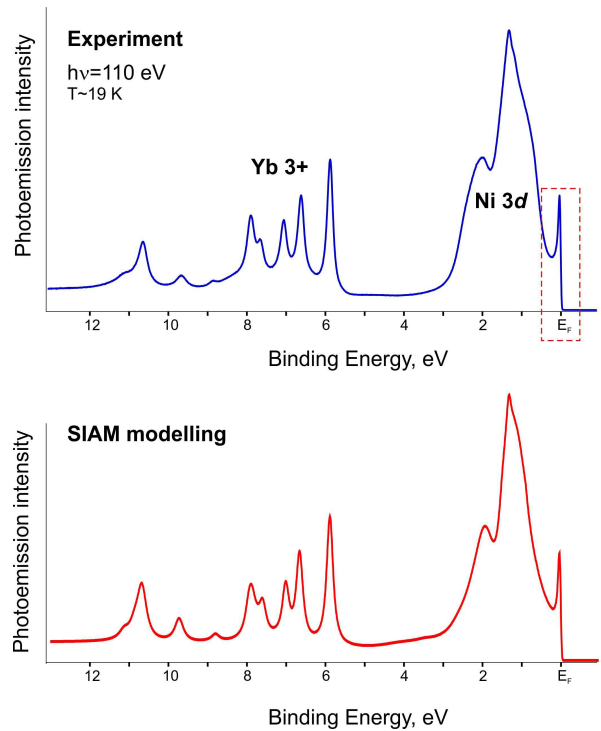}
\caption{(Color online) PE spectrum of the YbNiSn taken at $h \nu = 110\,$eV and $T = 19\,$K (upper panel) and its simulation by means of the
SIAM with hybridization strength $\Delta = 0.201\,$eV (lower panel).}
\label{pes}
\end{figure}

The upper panel of Fig.~\ref{pes} shows an angle integrated PE spectrum of YbNiSn, measured on a freshly cleaved sample, at photon energy $h \nu = 110\,$eV, where the relative intensity of $f$-emission is large. Clearly resolved features in the spectrum are: (i) the well-known broad $4f^{12}$ final state multiplet lying at a BE between 5.9 and 11.5~eV as it is expected from photoionization of a trivalent Yb $4f^{13}$ ground-state configuration, (ii) strong emissions from the Ni $3d$ valence band (VB) between 0.3 and 3~eV BE, and (iii) a sharp peak close to $E_F$ which was not predicted in the band-structure calculations. This feature is assigned to the $4f^{13}_{7/2}$ final states arising from a $4f^{14}$ admixture to the trivalent bulk ground state. Considering the SO splitting, the related $4f^{13}_{5/2}$ component is expected at $\sim$1.3~eV BE. Looking closely, a weak shoulder at the low-energy tail of the Ni $3d$ band could be seen and assigned to the respective state. The PE evidence for Kondo physics is the simultaneous observation of an intense Yb$^{3+}$ feature together with a sharp peak at $E_F$.

\begin{figure}
\begin{center}
  \includegraphics[width=0.90\columnwidth]{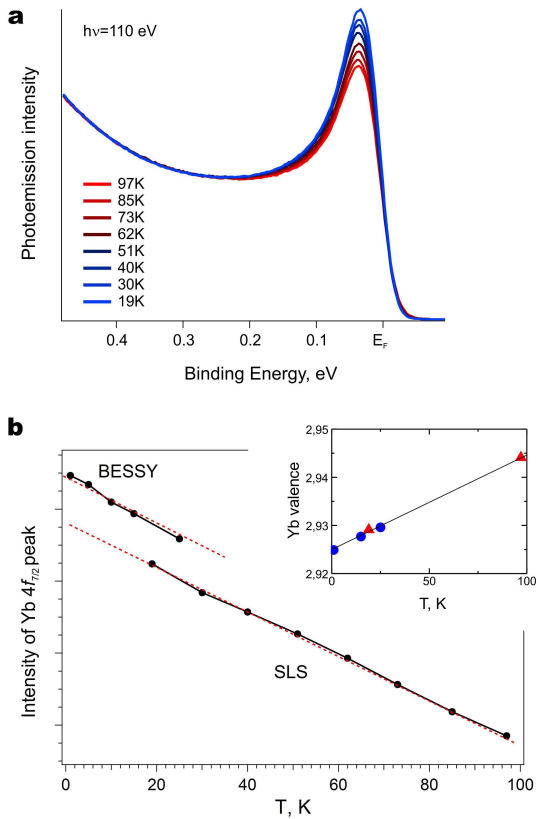}
\end{center}
\caption{(Color online) (a) PE peak near $E_F$, corresponding to the $4f^{13}_{7/2}$ final-state configuration, measured from 97~K to 19~K. (b) $T$-\,dependence of the integrated weight for two series of measurements. Inset: $T$-\,dependence of the Yb valence, obtained from the SIAM simulations based on the BESSY (blue circles) and SLS (red triangles) experiments.}
\label{peak_EF}
\end{figure}

This assignment implies that the resonance at $E_F$ is not perturbed by the surface valence instability towards Yb$^{2+}$. A respective divalent surface signal is expected between 0.5 and 1.0~eV BE ~\cite{Yb-val, tjeng93} and masked here by the intense Ni $3d$ emission. No measurable $T$-\,dependence is observed in this range, which allows to accurately normalize the raw PE spectra to the high BE tail of Ni $3d$ (Fig.~\ref{peak_EF}a) and attribute the $T$-\,dependence in the vicinity of $E_F$ only to the bulk excitations~\cite{patil14}.

The SIAM modeling in the lower panel of Fig.~\ref{pes} assumes an atomic-like $f$-shell (N$_f$=14) and a realistic band-structure input. A large SO interaction of 1.28~eV splits  the $4f^{13}$ final state into multiplets \textbf{\textit{J}}$=7/2$ and \textbf{\textit{J}}$=5/2$. The CEF splittings inside each multiplet are neglected. The simulation of the $T$-dependence (not shown) is phenomenological: A $T=0$ spectrum is first obtained by the Gunnarsson-Sch\"onhammer approach ~\cite{gunn,hayn01-prb} and a hybridization strength $\Delta(T)$ is then varied to model the evolution with $T$. By retaining N$_f$\,=14, a good overall fit up to the $4f^{12}$ final states is achieved. This allows to extract reliable estimates for $U$ and $n_f$ (for details see Supplemental Material ~\cite{SM}). 

 The nominal valence $\nu=2+n_f(T)$, as determined for five different temperatures, is shown in the inset of Fig.~\ref{peak_EF}b, confirming near trivalent Yb. A small deviation $n_f<1$ is already present at the highest $T$ and extends with a linear slope over at least an order of magnitude above and below the empirically defined $\overline{T}_K$.

Fig.~\ref{peak_EF}a shows the low energy part of the raw PE spectra, taken at SLS from 97~K down to 19~K. The peak intensity is clearly below  $E_F$, as expected for Yb.~\cite{tjeng93}  An influence of the Fermi function is noticeable only in the tail above $E_F$. Any fine structure, due to a difference between SIAM and PAM simulation, should show up here. The absence of marked fine structure in the case of YbNiSn can have several origins. The renormalized bands that form the coherent part of the $4f^{13}_{7/2}$ final state configuration may have smaller CEF splittings than in YbRh$_2$Si$_2$,~\cite{Zwicknagl} so that they all appear bunched together under the large resonance width of $\sim$50~meV. This can explain the absence of separate CEF satellites but does not exclude a fine structure due to lattice effects. We argue that there is a further convolution with an intrinsic broadening function, due to the particular difficulty of cleaving YbNiSn along well defined crystallographic planes.  

An exceptionally high peak intensity, in comparison with most other Yb compounds, is attributed to the low level of defects in the bulk. The surprising monotonic rise of the peak continued with decreasing $T$, as verified from 23~K to 1~K in a second experiment at BESSY-II. A typical characteristic of Kondo screening is a resonance that does not simply sharpen at constant weight. Rather, peak height and spectral weight increase together. A quantitative analysis of the weight, as obtained by integrating over the intensity, is plotted in Fig.~\ref{peak_EF}b. The spectra were normalized with respect to the VB emission for BE $>$0.25~eV. The $T$-\,dependence of the weight has the same linear slope in both data sets and it is remarkable that there is no measurable deviation at $T_C$. The extended interval of linear $T$-\,dependence, correlated with the evolution of $n_f(T)$ (inset), is the central experimental result, reported here.

\subsection{A toy model for PAM and its results} 

A PAM simulation with Kramers doublets only (N$_f$=2), using similar Anderson parameters as in the SIAM and a SE function $\Sigma_f(\omega,T)$ from DMFT,~\cite{Benlagra,notations,tahvildar,Meyer,Amaricci} leads to contradictory results for the $k$-integrated spectrum near $E_F$. On the one hand, extrapolating from the literature to our scenario, at the lowest available $T$, we find $\overline{T}_K^{(2)}\sim\,$0.72~meV$\equiv$8.4~K for the width of the QP band, in agreement with the empirical $\overline{T}_K$ from the bulk measurements. Self consistent fillings $n_f\sim\,$0.9 and $n_c\sim\,$0.2 confirm an asymmetric KL scenario, with $n=n_f+n_c=1.1$ slightly above "quarter filling", where competition with both FM and AFM order was found, depending on $U$ and $n$. 
A small number of holes in the VB, $n_c\rightarrow0$, is the regime of "protracted screening".~\cite{tahvildar,Arko} The main consequence for spectroscopy is a renormalized band width $\overline{T}_K^{(2)}<<T_K^{(2)}$, strongly reduced relative to the corresponding SIAM width.
 On the other hand, the model with Kramers doublets only leads to a $T$-\,dependence of the calculated QP band that disagrees strongly with experiment. The integrated weight drops initially $\propto (T/\overline{T}_K)^2$, followed asymptotically by log-linear decay.  

In our interpretation, both the large width of the PE resonance and the extended interval of linear $T$-\,dependence have a common origin in the effect of a set of closely spaced, already hybridized bands, resulting from the orbital degrees of freedom in the presence of rather weak CEF splittings. This implies that Kondo screening begins far above the empirical $\overline{T}_K$, a reasoning supported by the CEF effects in the SIAM. A large FS, present already at 100~K,~\cite{kummer2015} also fits into this picture and definitely demands a lattice model for its interpretation.

Our toy model is defined on the $k$-resolved level by a phenomenological replacement $\Sigma_f(\omega,T)\rightarrow\Sigma_{eff}(\omega,T)=\Sigma_{CFM}(\omega)+i\tilde{\Sigma}^{''}(T)$ of the SE from DMFT, otherwise keeping the same $2\times2$ Dyson equation.~\cite{notations}  The first term is obtained with the continued fraction method (CFM).~\cite{hayn,Chikina} The width of the four overlapping QP bands is increased to an effective KL scale $\overline{T}_K^{(8)}$. The CFM incorporates the Hubbard sum rules and the Luttinger sum rule, with FS volume $n/2$. Here, $n$ is kept constant as function of $T$ by self consistently varying the position of $\epsilon_c$.~\cite{tahvildar}  The self-consistency generates an implicit $T$-dependence in $\Sigma_{CFM}(\omega)$. The $\omega$-independent broadening $\tilde{\Sigma}^{''}(T)$ replaces $\Delta(T)$ in the SIAM simulation as fitting parameter. It has a more direct physical interpretation, since it models the uncertainty in the relative position of the four bands as a life time effect on the $k$-resolved level.~\cite{Smith} 

Results for the toy model, with Anderson parameters for an asymmetric KL scenario  (see Supplemental Material ~\cite{SM}), are shown in Fig.~\ref{PAM_simulation}. The effective damping rate, extracted from the measured $T$-\,dependence in Fig. \ref{peak_EF}b, is of a non-FL type: $\tilde{\Sigma}_f^{''}(T)=A+BT$, with a $T=0$ offset $A\sim2\overline{T}_K^{(8)}$ and a linear thermal slope $B=0.6$. 

\begin{figure}
\begin{center}
  \includegraphics[width=0.90\columnwidth]{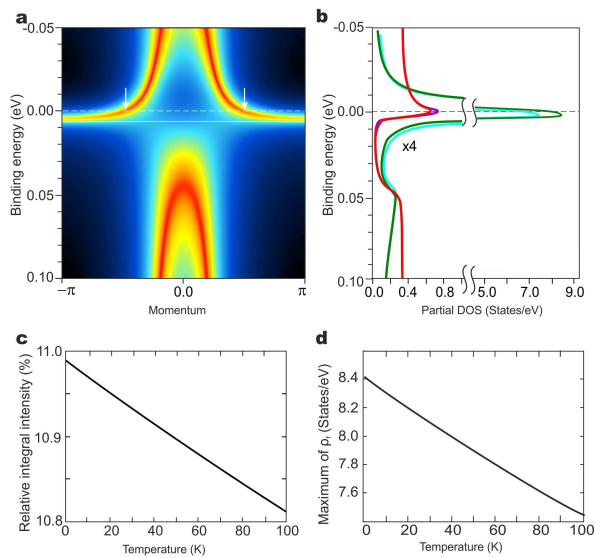}
\end{center}
\caption{(Color online) Fine structure around $E_F$, as simulated by the toy model. The half width of the bare VB is $W=$1~eV. The effective KL scale is $k_B\overline{T}_K^{(8)}=$5.3~meV. 
(a) \textit{k}-resolved $f$-intensity at $T=0$, showing a QP band and a less renormalized bonding band, both of finite width, repelling each other at the renormalized $f$-level $\epsilon_f^*$ (continuous line). The crossing points (arrows) lie on the large FS.  
(b) Partial DOS $\rho_f(\epsilon)$ (green line) and $\rho_c(\epsilon)$ (red line), at $T=$0~K and 100~K. The $f$-DOS dominates in an interval of $\pm$10~meV around $E_F$ and again around the top of the bonding band. Simulated $T$-\,dependences: (c) Weight of $\rho_f(\epsilon)$ (integrated over
$\pm$0.1~eV) as a percentage of the Pauli weight and (d) Maximum of $\rho_f$.}
\label{PAM_simulation}
\end{figure}

The $k$-resolved  plot of $f$-intensity at $T=0$, (Fig. \ref{PAM_simulation}a), shows a QP band and an anti-bonding band, both strongly renormalized and blurred by the anomalous damping. The $f$-intensity is high above and below the renormalized $f$-level $\epsilon_f^*$. For the definition of the FS, it is important that the term $\Sigma_{CFM}(\omega)$, in the absence of the phenomenological changes, can be mapped exactly on the DMFT+NRG solution in the limit $T\rightarrow0$, which has a discontinuity in the momentum distribution at $k_F$. This was checked by benchmarking with scenarios near quarter filling for $U = 3.5 W$ and $U = 10 W$.~\cite{Benlagra} In the presence of $\tilde{\Sigma}_f^{''}(0)$, the discontinuity is wiped out but a FS remains defined in ARPES  by the $k$-values where the intensity peaks at zero BE.~\cite{kummer2015}
The partial $f$-DOS $\rho_f(\epsilon)$ and valence band $c$-DOS $\rho_c(\epsilon)$ in Fig. \ref{PAM_simulation}b both show resonant behavior. The highest peak is near the bottom of the QP band. The linear $T$-dependence of both the weight and the peak height (Fig.~\ref{PAM_simulation}c) and (Fig.~\ref{PAM_simulation}d), in agreement with Fig.~\ref{peak_EF}, is explained by a $T$-\,linear damping inside the SE, on the $k$-resolved level.  

Note that $\epsilon_f^*$  is the center of gravity of the total split resonance. Here, its position close to the main peak is a signature of the asymmetric KL. The QP DOS alone, of width $\overline{T}_K^{(8)}$, does not correspond to the Kondo peak in the SIAM. When the split resonance is measured with insufficient resolution, the "Kondo peak" reappearing at $\epsilon_f^*$ includes both sides. We identify the overall width with the scale $T_K^{(8)}$ of the SIAM. The pseudogap can be wiped out intrinsically when the spacings between the CEF levels increase. For large spacings, each SO or CEF satellite is expected to be modified by a lattice signature. The $T$-dependence of the resonance weight is observable even with moderate resolution. A hierarchy of the various characteristic scales, evaluated for the case of YbNiSn, is presented in the Supplemental Material. ~\cite{SM}

\section{Summary}

In conclusion, PE spectra of the Kondo lattice compound YbNiSn reveal an unusual $T$-\,dependence of the Fermi-level peak, which includes contributions from the QP resonance and its unresolved CEF satellites. With decreasing temperature, the peak reveals a steady linear increase of intensity which extends over a large temperature range from 100~K to 1~K, without showing any peculiarities in the region of $\overline{T}_K$ $\sim$ $T_C$. 

In the light of the SIAM, these intensity variations reflect a linear increase of 4$f$ occupancy, indicating an onset of the Kondo lattice behavior at temperatures above 100~K. Within the PAM this phenomenon could be described by a non-Fermi liquid like $T$-\,linear damping in the self-energy which accounts phenomenologically for the feedback from closely spaced CEF-split bands. Given this interpretation, use of the "non-FL" terminology has to be taken with a grain of salt. In particular, we do not have enough evidence to ascribe the $T$-\,linear damping to the presence of a QCP at $T=0$.  

When combined with the large, $T$-invariant  FS in YbRh$_2$Si$_2$,~\cite{kummer2015}  our findings on the resonance near $E_F$ in the PE spectrum of YbNiSn  confirm that the formation of a coherent state at $\sim 10K$, as visible in the transport, has to be dissociated from the formation of renormalized bands and the beginning of Kondo screening at much higher $T$.

\section{Acknowledgments}

This work was supported by the German Research Foundation (DFG; grants VY64/1-3, GE602/2-1, GRK1621 and SFB1143) and by Research Grant No. 15.61.202.2015 of Saint Petersburg State University. DAS and ADH acknowledge support from EPSRC grant EP/J00099X/1. The authors would like to acknowledge Christoph Geibel for stimulating discussions and valuable suggestions.


\begin{thebibliography}{99}

\bibitem{gegenwart2008}
P.~Gegenwart, Q.~Si, and F.~Steglich, Nature Phys., \textbf{4}, 186 (2008).

\bibitem{Coleman} P. Coleman, Handbook of Magnetism and Advanced Magnetic Materials, Vol. {\bf1}, 95 - 148,  John Wiley and Sons, Ltd. (2007)
 
\bibitem{Anderson} P.W.Anderson, Phys. Rev. {\bf124}, 41 (1961).

\bibitem{NNH+92}
P.~Bonville et al., Physica B \textbf{182}, 105 (1992).

\bibitem{Cornut} 
B. Cornut and B. Coqblin, Phys. Rev. B \textbf{5}, 4541 (1972).

\bibitem{adroja98-physb}
D.T.~Adroja, B.D.~Rainford, T.~Takabatake, Physica B, \textbf{253}, 269 (1998).

\bibitem{Kroha} 
J. Kroha et al., Physica E (Amsterdam) \textbf{18}, 69 (2003).
\bibitem{kasaya92}
M.~Kasaya et al., J. Phys. Soc. Jpn. \textbf{60}, 3145 (1991).

\bibitem{klingner11-prb}
C.~Klingner et al., Phys. Rev. B \textbf{83}, 144405 (2011).

\bibitem{Benlagra} 
A. Benlagra, Th. Pruschke and M. Vojta, Phys. Rev. B {\bf 84}, 195141 (2011).

\bibitem{notations} Our notations for the PAM are as in ref. [\onlinecite{Benlagra}]. 

\bibitem{vyal10}
 D. V. Vyalikh et al., Phys. Rev. Lett. \textbf{105}, 237601 (2010).

\bibitem{Yb-val}
K.~Kummer et al., Phys. Rev. B \textbf{84}, 245114 (2011).

\bibitem{patil14}
S. Patil et al., JPS Conf. Proc. \textbf{3}, 011001 (2014).

\bibitem{kummer2015}
K.~Kummer et al., Phys. Rev. X \textbf{5}, 011028 (2015).


\bibitem{Allen} 
J. W. Allen et al., Adv. Phys. \textbf{35} 275 (1986).
J.W. Allen, J. Phys. Soc. Jpn \textbf{74}, 34Ð48 (2005).

\bibitem{patthey}
F. Patthey et al., Phys. Rev. Lett. \textbf{58} 2810 (1987).

\bibitem{reinert}
F. Reinert et al., Phys. Rev. Lett. \textbf{87} 106401 (2001).

\bibitem{ehm}
D. Ehm et al., Phys. Rev. B \textbf{76} 045117 (2007).

\bibitem{figure9} See the three panels of Fig.~9 in ref. [\onlinecite{ehm}].

\bibitem{tjeng93}
L. H. Tjeng et al., Phys. Rev. Lett. \textbf{71} 1419 (1993).

\bibitem{Zwicknagl} 
G. Zwicknagl, J. Phys. Cond. Matter {\bf 23}, 094215 (2011).
See the RB result for H=0 in Fig. 3.

\bibitem{Krasovskii} 
R. Kuzian and E. Krasovskii, Phys. Rev. B \textbf{94} 115119 (2016).

\bibitem{tahvildar}
A. N. Tahvildar-Zadeh, M. Jarrell, and J. K. Freericks, Phys. Rev. Lett. \textbf{80} 5168 (1998).

\bibitem{Pruschke} 
Th. Pruschke, R. Bulla, and M. Jarrell, Phys. Rev. B {\bf 61}, 12799 (2000).

\bibitem{Meyer} 
D. Meyer and W. Nolting, Phys. Rev. B \textbf{62}, 5657 (2000).

\bibitem{Amaricci} 
A. Amaricci et al., Phys. Rev. B {\bf 85}, 235110 (2012).

\bibitem{Kainz}  
A. Kainz et al., Phys. Rev. B {\bf 86}, 195110 (2012).

\bibitem{Costi} 
T. A. Costi, J. Kroha, and P. W\"olfle, Phys. Rev. B \textbf{53}, 1850 (1996).

\bibitem{Georges} A. Georges, L. De Medicis and J. Mravlje, Annual Review of Condensed Matter Physics {\bf4}, 137 (2013).
 
\bibitem{shim} J. H. Shim, K. Haule and G. Kotliar, Science \textbf{318}, 1615 (2007).

\bibitem{choi} H. C. Choi et al., Phys. Rev. Lett. {\bf108}, 016402 (2012).

\bibitem{shim1} J. H. Shim, private communication.

\bibitem{Joyce}
J. J. Joyce et al., Phys. Rev. Lett. \textbf{68}, 236 (1992).

\bibitem{Arko}  
A. J. Arko et al., J. Alloys and Compounds \textbf{271}, 826 (1998).

\bibitem{Lawrence}
J.M.~Lawrence et al., J. Magn. Mag. Mater. \textbf{108}, 215 (1992).

\bibitem{Weibel}
P.~Weibel et al., Z. Phys. B \textbf{91}, 337 (1993).

\bibitem{Blyth}
R.I.R.~Blyth et al., Phys. Rev. B \textbf{48}, 9497 (1993).

\bibitem{gunn}
O. Gunnarsson and K. Sch\"{o}nhammer,
Phys. Rev. B \textbf{31}, 4815 (1985).

\bibitem{hayn01-prb}
R.~Hayn et al., Phys. Rev. B \textbf{64}, 115106 (2001).

\bibitem{SM} See Supplemental Material on the theoretical modeling at
[URL will be inserted by publisher].

\bibitem{CeFePO} M.G.Holder et al., Phys. Rev. Lett. {\bf104}, 098402 (2010).

\bibitem{s4} M.G. Holder et al., Phys. Rev. B {\bf86}, 020506(R) (2012).
  
\bibitem{hayn}
R. Hayn, P. Lombardo, and K. Matho,
Phys. Rev. B \textbf{74}, 205124 (2006).

\bibitem{Chikina} 
A. Chikina, PhD thesis, TU Dresden (2016).

\bibitem{Smith} N. V. Smith, P. Thiry and Y. Petroff, Phys. Rev. B \textbf{47}, 15476 (1993).


\end{thebibliography}
\end{document}


\title{Supplementary Material}

\author{A. Generalov}
\affiliation{MAX-Laboratory, Lund University, Box 118, 22100 Lund, Sweden}

\author{D. A. Sokolov}
\affiliation{School of Physics and CSEC, Universuty of Edinburgh, Edinburgh EH9 3FD, UK}
\affiliation{Max-Planck-Institut f\"{u}r Chemische Physik fester Stoffe, D-01187 Dresden, Germany}

\author{A. Chikina}
\affiliation{Institut f\"{u}r Festk\"{o}rperphysik, Technische Universit\"{a}t Dresden, D-01062 Dresden, Germany}

\author{Yu. Kucherenko}
\affiliation{Institute for Metal Physics, National Academy of Science of Ukraine, UA-03142 Kiev, Ukraine}

\author{V. N. Antonov}
\affiliation{Institute for Metal Physics, National Academy of Science of Ukraine, UA-03142 Kiev, Ukraine}

\author{L. V. Bekenov}
\affiliation{Institute for Metal Physics, National Academy of Science of Ukraine, UA-03142 Kiev, Ukraine}

\author{S. Patil}
\affiliation{Department of Physics, Indian Institute of Technology, Banaras Hindu University, Varanasi-225001, India}

\author{A. D. Huxley}
\affiliation{School of Physics and CSEC, Universuty of Edinburgh, Edinburgh EH9 3FD, UK}

\author{J. W. Allen}
\affiliation{Randall Laboratory, University of Michigan, 450 Church St., Ann Arbor, MI 48109-1040, USA}

\author{K. Matho}
\affiliation{Institut N\'{e}el, C.N.R.S. and Universit\'{e} Grenoble Alpes, BP 166, F 38042 Grenoble cedex 9}

\author{K. Kummer}
\affiliation{European Synchrotron Radiation Facility, 71 Avenue des Martyrs, Grenoble, France}

\author{D. V. Vyalikh}
\affiliation{Institut f\"{u}r Festk\"{o}rperphysik, Technische Universit\"{a}t Dresden, D-01062 Dresden, Germany}
\affiliation{Donostia International Physics Center (DIPC), Departamento de Fisica de Materiales and CFM-MPC UPV/EHU, 20080 San Sebastian, Spain}
\affiliation{Saint Petersburg State University, Saint Petersburg 198504, Russia}
\affiliation{IKERBASQUE, Basque Foundation for Science, 48011 Bilbao, Spain}

\author{C. Laubschat}
\affiliation{Institut f\"{u}r Festk\"{o}rperphysik, Technische Universit\"{a}t Dresden, D-01062 Dresden, Germany}

\date{\today}

\maketitle

\section{SIAM simulation}
\label{SIAM}
The valence band (VB) density of states (DOS) for YbNiSn was determined by the fully relativistic spin-polarized version~\cite{NKA+83} of the LMTO method~\cite{And75} where the Yb $4f$ electrons were considered as quasi-core states. For SIAM calculations only those VB states were taken into account that are not forbidden to be hybridized with Yb $4f$ states, i.e. the local VB $f$ components at the Yb site that originate mainly from Ni $d$ and Sn $p$ states. The contribution of the Ni $3d$ emission to the PE intensity was taken to be proportional to the corresponding partial density of states.

In order to extract the mean Yb valence from the experimental data we have simulated the PE spectra on the basis of the SIAM. A variational solution may be obtained by a simple numerical procedure \cite{hayn01-prb} that represents a minimal version of the Gunnarsson-Sch\"onhammer approach \cite{gunn}. Taking into account the nearly filled 4$f$ shell of Yb, we used a hole representation of the code considering $h^0, h^1$ and $h^2$ configurations ($f^{14}$, $f^{13}$ and $f^{12}$, respectively) as basis functions. The main parameters of the model are the energy of the $h^1$ state $\epsilon_f$, the hopping parameter $\Delta$ and the on-site Coulomb repulsion $U_{ff}$. The code was additionally generalized to take into account multiplet effects.

The spectral function is obtained by projecting the hole state generated by the PE process onto the individual final states:

\begin{equation}\label{eqn:_SF_FS}
I(\epsilon) \sim -\mbox{Im} \sum_{f}
\frac{ | \langle f |\hat{T} | g \rangle |^2}
{h \nu-\epsilon-(E_f-E_g)+i\Gamma}.
\end{equation}

$| g \rangle$ denotes the ground state and the sum should be calculated over all final states $| f \rangle$ with one electron removed. The dipole transition operator $ \hat{T}$ in the matrix element depends in general on the photon energy $h \nu$. $\epsilon$ is the photoelectron kinetic energy and $\Gamma$ represents the spectral broadening due to finite lifetime of the excited final state. The parameter for the Lorentzian broadening was taken to be energy-dependent as $\Gamma = \Gamma_0 + \Gamma_1 E_B$ with $\Gamma_0 =0.03\,$eV and $\Gamma_1 =0.01$. The spectrometer resolution was taken into account by Gaussian broadening with $\Gamma_G =0.02\,$eV. With these procedures we have achieved an accurate description of experimentally observed widths for all the spectral lines.

The energies and intensities of the Yb $4f^{12}$ multiplet components were taken from Ref.\onlinecite{gerken83-jpfmp} where they were calculated for a free Yb ion. In order to achieve the best agreement with PE results obtained for a single crystal, we have enlarged the energy spacing between the multiplet components by the factor 1.115.

The SIAM code works nominally at $T=0$, where it captures the local Kondo resonance. Changes as function of $T$ are then described phenomenologically by a hybridization strength $\Delta(T)$ as fit parameter. Note that $n_f(T)$ depends essentially on the ratio $\Delta(T)/\epsilon_f$. The chosen value for $\epsilon_f$ leads to an optimal fit of the slightly asymmetric resonance shape at $E_F$. The main criteria for fitting were the integral intensity ratios of the $4f_{7/2}$ doublet component of the $4f^{13}$ configuration (at the Fermi level) to the leading $^3H_6$ component in the multiplet structure of the $4f^{12}$ configuration (near the binding energy of 6~eV).

Similar behavior of $n_f(T)$ was reported for the heavy-fermion systems YbRh$_2$Si$_2$~\cite{kummer2015} and Yb$_2$Pd$_2$Sn,~\cite{yamaoka2012} for $T>T_K$.  For YbRh$_2$Si$_2$, a saturation was observed below $T_K$.~\cite{kummer2015}  A stronger slope was observed in the mixed-valent compound YbNi$_3$Al$_9$.~\cite{Utsumi} No $T$-dependence was reported for trivalent compounds YbPd$_2$Sn~\cite{yamaoka2012} and YbNi$_3$Ga$_9$.~\cite{Matsubayashi} 

\section{PAM simulation}

\label{PAM}
We use the $2\times2$ matrix of  $k$-resolved Green functions for the PAM, given e.g. by eqs. (B6) - (B8) in Ref.~\onlinecite{Benlagra}.~\cite{notations} With the replacement
\begin{equation}
\Sigma_f(\omega,T)\rightarrow\Sigma_{eff}(\omega,T)=\Sigma_{CFM}(\omega)+i\tilde{\Sigma}^{''}(T)
\label{Sigeff}
\end{equation}
we are leaving the microscopic framework of DMFT.~\cite{Benlagra,tahvildar,CeFePO,s4,Amaricci} To keep the $2\times2$ Dyson matrix, i.e. a two-band model, is the minimal framework that allows to study the competition between local repulsion $U$ and hybridization $V$ in a lattice environment.

In sect.~(\ref{Toy}), supplementary information is given on $\Sigma_{eff}(\omega,T)$, concerning the following issues. (i) Definition of a complete set of Anderson parameters and further input, needed to specify a thermodynamic equilibrium state without broken symmetry. (ii) The method of incorporating the Luttinger sum rule as a phenomenological constraint. (iii) Definition of $\Sigma_{CFM}(\omega)$ and of the parameter $Z_{eff}$ that allows to tune the low energy scale $\overline{T}_K$. (iv) The  self consistency routine. (v) Critical comments. 

Sect.~(\ref{Numerics}) contains details on the application to YbNiSn.

\subsubsection{Self energy for the toy model}
\label{Toy}
(i) The function $\Sigma_{eff}(\omega,T)$ describes a PAM, from which the higher $\textbf{\textit{J}}$-multiplet under the SO splitting has been projected out. 
To keep in line with DMFT, a possible $k$-dependence is neglected. The phenomenological tuning parameter is an increased degeneracy N$_f>2$. In the application to Yb, the ground multiplet $\textbf{\textit{J}}=7/2$ has degeneracy N$_f$=8. The focus is on a simulation of the low energy resonance at $E_F$ that appears near the top of the VB. 
 
With CEF splittings set to zero, the model input $[U, V, W, \epsilon_f, \epsilon_c]$ is completed by specifying the half width $W$ of the VB, as well as the positions relative to $E_F$ of the bare $f$-level, $\epsilon_f$, and the VB center, $\epsilon_c$. 

The connection to the single-impurity Kondo model, as well as to the KL model is established by the Schrieffer-Wolff expression for the dimensionless exchange coupling
\begin{equation}
g=\frac{\Gamma_0U}{\epsilon_f(\epsilon_f+U)}<0.
\label{coupling}
\end{equation}  
Here, $\Gamma_0$ is the Anderson-Friedel width, expressing the bare hybridization strength at $k_F$. Taking only the leading, exponential dependence on the orbital multiplicity into account,~\cite{Cornut} an enhancement of the Kondo temperature for N$_f>2$ over that for a Kramers doublet can be roughly estimated as
\begin{equation}
\frac{T_K^{(N_f)}}{T_K^{(2)}}=exp(\frac{2-N_f}{2gN_f})>1.
\label{enhancement}
\end{equation}
This relation for the SIAM is used in the toy model also as an estimate for the ratio $\overline{T}_K^{(N_f)}/\overline{T}_K^{(2)}$. For fixed N$_f$, the ratio $T_K^{(N_f)}/\overline{T}_K^{(N_f)}$ then depends strongly on $n_c$, the number of particles or holes in the VB. This relationship was so far explored microscopically only for N$_f$=2.~\cite{tahvildar}  

The evolution as function of the thermodynamic variables $T$ and $n$ is modeled, as usual, at fixed distance of the $f$-level from the VB center, $E_f\equiv\epsilon_f-\epsilon_c$. Keeping both $\epsilon_f$ and $\epsilon_c$ constant,~\cite{Benlagra} is equivalent to fixing the Fermi edge relative to the band center, resulting in a self-consistent variation of the particle number $n(T,E_F)$. Our  strategy is to simulate the $T$-dependence at fixed particle number, with self-consistent variation of the Fermi edge $E_F(T,n)\equiv-\epsilon_c$.~\cite{tahvildar} For an Yb-scenario, it is convenient to define $n=n_c+n_f$ as the number of holes per lattice site.

(ii) The FS of the KL in a ground state without broken symmetry is defined  in terms of a discontinuity in the momentum distribution ~\cite{KK} at $T=0$. The Luttinger sum rule states that the FS volume is proportional to $n=n_f+n_c$. In the presence of an ordered phase below $T_C$, this "large" FS appears at $T>T_C$, suitably broadened by the thermal fluctuations. 

Neglecting momentum dependence everywhere, except in the bare band structure $E_c(k)=\epsilon_k-\epsilon_c$, entails a number of simplifications which the toy model has in common with DMFT. In particular, the Luttinger FS is predetermined in shape and volume by the bare VB input. It is characterized by the kinetic energy at $k_F$, $E_c(k_F)\equiv E_{cF}$, which depends only on $n$ and monotonically sweeps the interval $[-W<E_{cF}(n)<W]$. The position of the kinetic energy relative to the Fermi edge, $\epsilon_F\equiv E_{cF}+\epsilon_c$ is usually of order $W$, i.e. not a strongly renormalized quantity. It plays nevertheless an important role in the self-consistent strong coupling solution of the PAM. Its occurrence in the exact relations listed below reflects the fact that the renormalized bands in the PAM are a superposition of coherent waves of both $f$- and $c$-character (see point (iii)). The number of $k$-points on the FS is related to the bare VB-DOS, $\rho_c^{(0)}(\epsilon_F)$. The Anderson-Friedel width is $\Gamma_0=V^2\rho_c^{(0)}(\epsilon_F)$.  

(iii) The continued fraction method (CFM),~\cite{hayn} combining the Hubbard-1 moment sum rules ~\cite{Hubbard} with the Luttinger sum rule ~\cite{Luttinger} and a Hermiticity condition $\Im{\Sigma_{CFM}}(0)=0$,~\cite{Martin} was recently generalized to the PAM.~\cite{Chikina} The previous application of $\Sigma_{CFM}(\omega)$ was restricted to Kramers doublets only and to $T=0$. It is mainly useful to get an analytical representation of DMFT results for $\Sigma_f(\omega,T\rightarrow0)$ ~\cite{CeFePO,s4} that are known only numerically. This permits to obtain a quick overview of the PAM spectra in the presence of a large FS, for scenarios with large $U$ and small $V$, as needed for a realistic modeling of RE compounds. The anaytic $\Sigma_{CFM}(\omega)$  covers KL and mixed valence regimes, including the dependence on asymmetric band fillings far from $n_c\sim1$.

Given the sum rules, the CFM ansatz has only a single tuning parameter
\begin{equation}
\frac{1}{Z_{eff}}=1-\frac{d\Re{\Sigma_{CFM}(\omega)}}{d\omega} |_{\omega=0} =1-\alpha
\label{Zeff}
\end{equation}
that can formally be varied in the interval $0<Z_{eff}<1-n_f/2$. The value belonging to the DMFT solution, $Z_{eff}=Z$, as extracted numerically from the slope $\alpha$ via eq.~(B25), is always in the allowed range. The self-consistency routine for $\Sigma_{CFM}(\omega)$ is extremely fast, compared to NRG, and the analytic solution is free of any artificial broadening, introduced in the NRG for convergence. Recent benchmarking near quarter filling, using two large $U$ scenarios solved with DMFT+NRG,~\cite{Benlagra} has confirmed good agreement over the entire range of frequencies.~\cite{Chikina} The successful benchmarking is also a check for the DMFT solution, by explicitly demonstrating that the NRG solver obeys the Luttinger sum rule. The important low energy scale, determining the width of the QP band, is $\overline{T}_K=Z\Gamma_0$,~\cite{notations} noted now as $\overline{T}_K^{(2)}$.  

Before discussing the use made of $\Sigma_{CFM}(\omega)$ in the toy model, at finite $T$ and with a rescaled $Z_{eff}$, we list a number of exact relations, valid in the absence of the damping function $\tilde{\Sigma}^{''}(T)$, in particular for the DMFT solution in the limit $T\rightarrow0$. Some of these relations are stated for the first time, others correct diverging statements, made in the literature.

In the Hubbard model, the renormalization constant $Z$, defined in analogy to eq. (\ref{Zeff}) by the slope $\alpha$, has two simultaneous interpretations, namely as the QP weight and as the dimensionless reduction factor of the Fermi velocity. They coincide only in the case of a one band model

In the PAM, $Z$ is the sum over the two coherent residues in the Green function of $f$-character, such as reminiscent in the two broadened peaks at constant $k$ in Fig. 5a. In all renormalized quantities involving both f- and c-character, the kinetic energy $\epsilon_F$, defined above, plays an important role.  The reduction factor of the Fermi velocity (i.e.: the inverse of the mass enhancement) is $v^*=ZV^2/(ZV^2+\epsilon_F^2)$. The QP weight at $k_F$, $Z_{QP}=v^*+Z(1-v^*)$ is always slightly larger than $v^*$. The renormalized $f$-level is $\epsilon_f^*=ZV^2/\epsilon_F$. 

The trajectory of the QP pole in the complex $\omega$-plane is $\omega_{QP}\approx v^*q+iZ\beta (1-v^*)(v^*q)^2$; $q=\epsilon_k-\epsilon_F$ ($|q|<<W$). The parameter $\beta$ is related to the initial curvature of the self energy (SE). This generic Fermi liquid (FL) behavior is found for large $U$. Non-FL deviations were found for moderate  $U\sim2W$, matching the band width.  These were analyzed near quarter filling in terms of a competition with AFM order, possibly replaced by FM order already around $n=1.1$.~\cite{Amaricci} For large $U$, the phase diagram is unknown.

Finally, as long as the SE is local, there are important "no-go" theorems for the partial DOS at the Fermi level, also called unitarity limits: $\rho_c(\epsilon=0,T=0)=\rho_c^{(0)}(\epsilon_F)$ and $\rho_f(\epsilon=0,T=0)=(\epsilon_F/V)^2\rho_c^{(0)}(\epsilon_F)$. Besides on bare model input, they depend only on the self-consistent $\epsilon_F$ and on the FL condition $\Im{\Sigma_{CFM}}(0)=0$. When the latter is dropped, they still represent valuable upper bounds, as in Fig. 5b.

(iv) The term $\Sigma_{CFM}(\omega)$ in eq. (\ref{Sigeff}) has a  low energy sector with a rescaled $Z_{eff}$, fixed by the relation $\overline{T}_K^{(N_f)}/\overline{T}_K^{(2)}=Z_{eff}/Z=exp[(2-N_f)/2gN_f]$, using eqs. (\ref{coupling}) and (\ref{enhancement}). The anomalous damping removes the FL condition, replacing it by $\Im{\Sigma_{eff}}(\omega=0,T=0)=\tilde{\Sigma}^{''}(T=0)$. 

Using this ansatz, with fixed $n$ and $Z_{eff}$, maintains an invariant FS as function of $T$ in the toy model. The trace over partial DOS of $f$- and $c$-character, calculated with $\Sigma_{eff}(\omega,T)$, has to obey the self-consistency condition
\begin{equation}
n\equiv n_f(T)+n_c(T)=2\Sigma_{i=f,c}\int_{-\infty}^{\infty} d\epsilon \rho_i(\epsilon,T)\frac{exp(\epsilon/T)}{1+exp(\epsilon/T)}. 
\label{ni}
\end{equation}

In fact, since the Hubbard-1 sum rules in the PAM depend on $n_f$, rather than on the total $n$, each term in the sum has its own self-consistency loop. A unique solution for $n_f(T)$ and $n_c(T)=n-n_f(T)$ was found in all simulations. This feedback of $n_f(T)$ causes an implicit $T$-dependence also in the term $\Sigma_{CFM}(\omega)$.
 
(v) A critical comment about the toy model is in order. The $2\times2$ matrix of $k$-resolved Green functions, eqs. (B6) - (B8), is independent of the orbital index $\sigma$ inside a Kramers doublet ("spin s=1/2" in Doniach's KL model). A generalization to N$_f$ identical blocks, independent of the orbital colors, is formally possible in a so called "spherical cow" model,~\cite{Coleman} in which the $f$-shell has SU(N$_f$) symmetry and there are N$_f$ overlapping FS sheets, each with enclosed volume $(n_f+n_c)/N_f$. This does not correspond to the observed electronic structure. The toy model has a rescaled low energy sector, taking N$_f>2$ into account.  But the FS volume is $n/2$, which has to be interpreted in such a way, that only one Kramers degenerate band among the four closely spaced bands actually disperses across the Fermi edge. It is expected that a microscopic solution for a PAM with small CEF splittings, necessitating a full $N_f\times N_f$ Dyson matrix, will be able to lift this inconsistency. 

\subsubsection{Application to YbNiSn}
\label{Numerics}

A $T=0$ scenario close to the SIAM simulation is $[U=6, V=0.25, W=1, \epsilon_f=-0.22, \epsilon_c=0.78]$. With $W=1 eV$ for the half width, energies are now all in eV, $\epsilon_f$ and $\epsilon_c$ are expressed as BE's. For simplicity, $\epsilon_f=\epsilon_c-1$ is held fixed at the top of $\rho_c^{(0)}(\epsilon)$. The self-consistent hole densities are $n=1.1$, $n_f=0.9$, $n_c=0.2$.

The estimated width of the QP band $\overline{T}_K^{(2)}=Z\Gamma_0\sim\, 0.72 meV\equiv 8.4 K$, reported in the main text, corresponds to $Z\sim0.012$. This $Z$-value is obtained by extrapolating numerical data, existing in the literature.~\cite{Benlagra,Amaricci} Note that $Z$ vanishes linearly with $n\rightarrow1$.~\cite{Amaricci} The main discrepancy of a model with Kramers doublets only, in comparison to the PE data, is a fast decay of the calculated resonance peak on tis temperature scale $\overline{T}_K^{(2)}\sim$ 8.4 K. 

With a dimensionless coupling constant $g=-0.188$, as calculated for this scenario from eq. (\ref{coupling}), an increased value $Z_{eff}=0.088$ is obtained from the enhancement factor in eq. (\ref{enhancement}). The associated $\overline{T}_K^{(8)}=Z_{eff}V^2\sim$ 5.3 meV $\equiv$ 61.5 K is also the scale of the first Kondo like regime with $d\rho/dT<0$, present in Fig. 2. 

For $n=1.1$, the kinetic energy on the FS is $E_c(k_F)=-0.05$, near the center of the VB. When expressed as a BE, the numerical value $\epsilon_F(T=0)=0.83$ yields $\epsilon_f^*=Z_{eff}V^2/\epsilon_F\sim$ 6.4 meV for the renormalized $f$-level. The continuous white line in Fig. 5a, representing $\epsilon_f^*$, is also the center of gravity of the total split resonance. The width of the split resonance is about an order of magnitude larger than the QP width. This is due to the asymmetry, that increases as the number of $c$-holes decreases. The corresponding SIAM scale is $T_K^{(8)}\sim$ 50 meV, which corresponds to the observed PE width.
It is noteworthy that the further enhancement of Kondo screening energy by a hypothetical f-shell with multiplicity N$_f$=14, according to eq. (\ref{enhancement}), is only $T_K^{(14)}\sim$ 66 meV. In the presence of a SO splitting of 1.28 eV, the screening of this multiplet cannot develop and it is safe to project out the excited multiplet $\textbf{\textit{J}}=5/2$ when the resonance at $E_F$ is modeled. As the SIAM simulation shows, both multiplets are necessary to model the two-hole final states.

The broadening $\tilde{\Sigma}^{''}(T)$ acts on the $k$-resolved level. A $T$-linear ansatz, with optimal parameters as reported in the main text, explains the variation of the integrated weight and the peak height of the DOS, plotted in Figs. 5c and 5d. The damping also leads to a weak linear $T$-dependence of $n_f(T)$, as well as parallel shifts of $\epsilon_F$ and $\epsilon_c$. These reveal a small variation of $E_F$,  relative to the center of the VB.

The damping in the calculated spectrum (Fig. 5b) is not strong enough to wipe out the pseudo gap. Possible reasons that this signature of translational invariance is not observed in the PE spectrum are discussed in the main text. 

The unitarity limits are $\rho_c(\epsilon=0,T=0)=\rho_c^{(0)}(\epsilon_F)\sim0.64$ and $\rho_f(\epsilon=0,T=0)=(\epsilon_F/V)^2\rho_c^{(0)}(\epsilon_F)\sim7.0$. In spite of the non-FL damping, already present at $T=0$, these upper bounds are almost reached in Fig. 5b. The shape of the narrow resonance, straddling $E_F$, depends on details of the input for the bare $\rho_c^{(0)}(\epsilon)$. This is an important difference to the universal shape of the Kondo peak in the SIAM. In the square lattice near quarter filling, the peak coincides with the lower band edge and shoots well above the unitarity limit.~\cite{Benlagra} This is due to the step like van Hove singularity, generic for a 2D band edge. In the YbNiSn simulation we used the semi-elliptic $\rho_c^{(0)}(\epsilon)$,~\cite{Amaricci} because its band edge is generic for 3D. The asymmetric peak in Fig. 5d overshoots the unitarity limit. When the toy model simulation is repeated for the $\rho_c^{(0)}(\epsilon)$ of the square lattice,~\cite{Benlagra} van Hove singularities are almost wiped out by the non-FL damping.